\documentclass[prx,reprint,superscriptaddress,twocolumn,showkeys]{revtex4-1}
\usepackage[utf8]{inputenc} 
\usepackage{amsmath}
\usepackage{braket}
\usepackage{graphicx}
\usepackage{pgfplots}
\usepackage{csquotes}
\usepackage{hhline}
\usepackage{amssymb}

\usepackage{dcolumn}
\usepackage{tabularx}
\usepackage[colorlinks=true,linkcolor=blue,citecolor=blue,urlcolor=blue]{hyperref}
\usepackage{longtable}
\usepackage{braket}
\usepackage{float}
\usepackage{listings}
\usepackage{color}
\definecolor{mygreen}{rgb}{0,0.6,0}
\definecolor{mygray}{rgb}{0.5,0.5,0.5}
\definecolor{mymauve}{rgb}{0.58,0,0.82}

\newcolumntype{C}{>{\centering\arraybackslash}X}
\begin{document}

\title{A Simulational Model for Witnessing Quantum Effects of Gravity Using IBM Quantum Computer}

\author{Manabputra}
\email{manabputra@gmail.com}
\affiliation{School of Physical Sciences,\\ National Institute of Science Education and Research, HBNI, Jatni 752050, Odisha, India}

\author{Bikash K. Behera}
\email{bkb13ms061@iiserkol.ac.in}
\author{Prasanta K. Panigrahi}
\email{pprasanta@iiserkol.ac.in}
\affiliation{Department of Physical Sciences,\\ Indian Institute of Science Education and Research Kolkata, Mohanpur 741246, West Bengal, India}

\begin{abstract}
Witnessing quantum effects in the gravitational field is found to be exceptionally difficult in practice due to lack of empirical evidence. Hence, a debate is going on among physicists whether gravity has a quantum domain or not. There had been no successful experiments at all to show the quantum nature of gravity till two recent independent works by Bose \emph{et al.} [Phys. Rev. Lett. \textbf{119}, 240401 (2017)] and by Marletto and Vedral [Phy. Rev. Lett. \textbf{119}, 240402 (2017)]. The authors have proposed schemes to test the quantumness of gravity in two small test masses by entangling two spatially separated objects using gravitational interactions. They provide a method to witness the entanglement using spin correlation measurements, which could imply evidence for gravity being a quantum coherent mediator. Here we propose a simulational model by providing a new quantum circuit for verifying the above schemes. We simulate the schemes for the first time in IBM's 5-qubit quantum chip `ibmqx4' by developing a quantum system which shows effects analogous to quantum gravity and calculates the degree of entanglement of the spin correlation. The entanglement witness over a range is obtained for different experimental parameters. 
\end{abstract}


\maketitle

{\em Introduction.-} The quantum theory and general relativity claim to be universally applicable with high accuracy in their respective domains. However, merging them into a unique theory so called quantum gravity had been a great challenge in contemporary physics \cite{qgs_SinghCurrSci2015,qgs_KhrennikovFoundPhys2017}. Though, physicists have already made a tremendous effort to achieve this goal \cite{qgs_WittGAQFT2003,qgs_KibbleCMP1979,qgs_PolchinskiIBS2005}, due to critical challenges and conceptual difficulties, it has been in vain \cite{qgs_KieferIMP2013,qgs_IshamICGRG1995,qgs_BorzeszkowskiTMQG1988}. Witnessing quantum effects like entanglement in the domain of gravity has been extremely difficult due to very weak gravitational interaction. Even in the past, it has been claimed that detection of graviton is impossible to realize physically \cite{qgs_DysonIJMP2012,qgs_BoughnCGG2006}. Hence, the lack of experimental detection of quantisation of gravity has led the scientific community question whether gravity has a quantum domain \cite{qgs_KieferAP2006,qgs_RothmanFP2006,qgs_OritiAQG2009}.              

The interaction of gravity has been observed at smaller distances and small time scales \cite{qgs_BiswasPRL2012}, hence opening a challenge to test whether gravity can be considered as a quantum entity or not. A number of cosmological models have been attempted, however, it has not been confirmed yet with conclusive experimental evidence \cite{qgs_HoseenfelderCQG2011,qgs_AshoorioonPLA2014,qgs_PikovskiNatPhys2012,qgs_AlbrechtPRA2014}. Physicists even have argued that gravity may not be a fundamental force \cite{qgs_ShandarinPRL1995,qgs_SakharovGRG2000} and its quantisation is not needed \cite{qgs_PenroseGRG1996}. According to Ronsenfeld, introducing gravitation into a general quantum field theory is an open problem due to the lack of empirical evidence though it is not difficult to develop mathematical model of quantum formalism for gravitation \cite{qgs_RosenfeldQGE1966}.        

For illustrating quantum nature of gravity, Bronstein \cite{qgs_GorelikPU2005,qgs_BronsteinPZS1936} and Feynman \cite{qgs_FeynmanCHCP1957} have taken initiatives to propose experimental schemes. The idea of preparing a test mass in superposition of two different locations and then interacting with the gravitational field \cite{qgs_BahramiarXiv2015,qgs_PagePRL1981,qgs_AnastopoulosCGG2015,qgs_DerakhshaniJPCS2016,qgs_DerakhshaniJPCS2016,qgs_DerakhshaniarXiv2016,qgs_MarlettoNat2017,qgs_MarlettoarXiv2018} has been started to take hold. The thought experiment proposed by Feynman required a full interference of the mass that could stop applying at a certain scale which consequently leads to the discovery of a new law of nature \cite{qgs_DiosiPLA1987,qgs_KharolihazyNC1966}. The proposed interference was not enough to demonstrate the quantum effect of the gravitational field, as it could be performed by a classical gravitational field such as Collela-Overhauser-Werner (COW) experiment \cite{qgs_ColellaPRL1975}. Some other prominent examples have also been encountered where gravitational redshift caused by Earth's gravitational field which is completely classical in nature \cite{qgs_MargalitSci2015,qgs_PikovskiNatPhys2015,qgs_AhluwaliaPRD1998}. From the above observation, it can be concluded that gravitationally induced phase in a quantum state does not imply the quantisation of gravity. Rather it is needed to verify that the gravitational field exists in a quantum superposition of two different values.              
Though there are many theoretical models of quantum gravity, none have been able to give any empirical evidence so far. Due to the weakness of gravity, it is extremely hard to study gravitational interactions between two test masses directly in any laboratory environment. This is where the use of quantum entanglement \cite{qgs_HorodeckiRMP2009} comes into play. Entanglement is a very unique quantum phenomenon. Once entangled, systems can no longer be written as a factorized tensor product of individual quantum states. This essentially means that entangled systems share a certain type of correlation that can be witnessed, even if the states are spatially separated \cite{qgs_SchrodingerPCPS1935}. Bose \emph{et al.} \cite{qgs_BosePRL2017} and Marletto and Vedral \cite{qgs_MarlettoPRL2017} have designed similar experiments based on the principle that two entangled objects always need a quantum mediator. Their proposal provides a much simpler yet elegant methodology for witnessing quantum nature of gravity. The scheme proposed by Marletto and Vedral \cite{qgs_MarlettoarXiv2018} gives a proof-of-principle for a field to have quantum effect. It states that if two quantum systems each in superposition of two different places become entangled through an interaction with a third system, then the third system is said to be a quantum entity. 

{\em Scheme.-} The scheme proposed by Bose \emph{et al.} \cite{qgs_BosePRL2017} requires two mesoscopic test masses of mass $m_1$ and $m_2$. A quantum mechanical phase $\frac{E\tau}{\hbar}$ is being induced, where E=$\frac{Gm_1m_2}{d}$; $d$ is the spatial separation between the two test masses and $\tau$ is the interaction time, to the system which suffices for gravitational interaction entangling the two masses. It is assumed that the gravitational interaction is mediated by the gravitational field and that entanglement cannot be created using local operations and classical communication (LOCC) \cite{qgs_BosePRL2017,qgs_HorodeckiRMP2009}. Thus only the mutual gravitational interaction entangles the state implying that the mediating gravitational field has to be quantised. The state evolution is given by the following expressions.

\begin{align}
|\Psi(t=0)\rangle_{12} =\frac{1}{\sqrt{2}}( |0\rangle_1+ |1\rangle_1)\frac{1}{\sqrt{2}}( |0\rangle_2+ |1\rangle_2)\nonumber\\ 
=\frac{1}{2}( |00\rangle_{12}+ |01\rangle_{12}+|10\rangle_{12}+ |11\rangle_{12})\nonumber\\
\rightarrow |\Psi(t=\tau)\rangle_{12} =\frac{1}{2}(e^{i\phi}|00\rangle_{12}+e^{i\phi_{LR}}|01\rangle_{12}\\+e^{i\phi_{RL}} |10\rangle_{12}+ e^{i\phi}|11\rangle_{12})\nonumber \\
=\frac{e^{i\phi}}{\sqrt{2}}\{|0\rangle_1\frac{1}{\sqrt{2}}( |1\rangle_2+e^{i\Delta\phi_{LR}}|1\rangle_2)\nonumber\\+|1\rangle_1\frac{1}{\sqrt{2}}(e^{i\Delta\phi_{RL}} |0\rangle_2+ |1\rangle_2)\}
\end{align}

where,  $\Delta\phi_{RL} =\phi_{RL} -\phi, \Delta\phi_{LR}=\phi_{LR}-\phi$  and \\

$\phi_{RL} \sim  \frac{G m_1 m_2 \tau}{\hbar (d-\Delta x)},\,\,\,\phi_{LR} \sim  \frac{G m_1 m_2 \tau}{\hbar (d+\Delta x)},\,\,\, \phi \sim \frac{G m_1 m_2 \tau}{\hbar d}$\\

are the gravitational phases and $|0\rangle$ and $|1\rangle$ are the two spatially separated states for the masses and $\Delta x$ is the separation between the centres of $|0\rangle$ and $|1\rangle$.\\ 
Clearly, $|\Psi(t=\tau)\rangle_{12}$ is not factorizable and hence is an entangled state of the two ``orbital qubits" $\frac{1}{\sqrt{2}}( |L\rangle_2+e^{i\Delta\phi_{LR}}|R\rangle_2)$ and $\frac{1}{\sqrt{2}}(e^{i\Delta\phi_{RL}} |L\rangle_2+ |R\rangle_2)$ where the entanglement is created due to the gravitational interactions only (neglecting the short-range Casimir-Polder force \cite{qgs_CasimirPR1948}).\\
Stern-Gerlach Interferometry is used on the test masses with embedded spins to create a superposition of spatially separated state and hence the orbital entanglement is naturally mapped to the spin entanglement. The spin entanglement witness is then measured in two complementary  bases.\\
Let $|\tilde{0}\rangle$ and $|\tilde{1}\rangle$ corresponds to the up and down spin state respectively and the centre of mass state of the test mass $m_i$ is $|C\rangle_i=\frac{1}{\sqrt{2}}(|0\rangle_i+ |1\rangle_i)$.  \\
So, a spin dependent spatial splitting of $|C\rangle$ in a non homogeneous magnetic field obeys the state evolution--\\
$|C\rangle_i \frac{1}{\sqrt{2}}
(|\tilde{0}\rangle_i+ |\tilde{1}\rangle_i)\rightarrow\frac{1}{\sqrt{2}}(|0\tilde{0}\rangle_i+ |1\tilde{1}\rangle_i)$\\
Then this coherent state is held for time $\tau$, where $\tau$ is the coherence time, by switching off the magnetic field of the Stern-Gerlach apparatus for time $\tau$.

The final step is to bring back the superposition using unitary Hadamard transformation of the mass state by refocusing the magnetic field in the opposite direction.
$|0\tilde{0}\rangle_i \rightarrow |C,\tilde{0}\rangle_i, |0,\tilde{0}\rangle_i \rightarrow |C,\tilde{0}\rangle_i$\\
The overall state evolution of this mass-spin system is therefore--
\begin{align}
    |\Psi(t=0)\rangle_{12} =\frac{1}{\sqrt{2}}( |0\rangle_1+ |1\rangle_1)\frac{1}{\sqrt{2}}( |0\rangle_2+ |1\rangle_2)|\tilde{0}\tilde{0}\rangle_{12}\nonumber\\
    =\frac{1}{2}( |00\tilde{0}\tilde{0}\rangle_{1212}+ |01\tilde{0}\tilde{0}\rangle_{1212}+|10\tilde{0}\tilde{0}\rangle_{1212}+ |11\tilde{0}\tilde{0}\rangle_{1212})
\end{align}

\begin{align}
\rightarrow|\Psi(t=t_{\text{End}})\rangle_{12} =\frac{1}{2}|C\rangle_1|C\rangle_2(e^{i\phi}|\tilde{0}\tilde{0}\rangle_{12}+e^{i\phi_{LR}}|\tilde{0}\tilde{1}\rangle_{12}\nonumber\\
+e^{i\phi_{RL}} |\tilde{1}\tilde{0}\rangle_{12}+ e^{i\phi}|\tilde{1}\tilde{1}\rangle_{12})\nonumber\\
=\frac{e^{i\phi}}{\sqrt{2}}.|C\rangle_1|C\rangle_2\{|\tilde{0}\rangle_1\frac{1}{\sqrt{2}}( |\tilde{1}\rangle_2+e^{i\Delta\phi_{LR}}|\tilde{1}\rangle_2)+\nonumber\\
|\tilde{1}\rangle_1\frac{1}{\sqrt{2}}(e^{i\Delta\phi_{RL}} |\tilde{0}\rangle_2+ |\tilde{1}\rangle_2)\}
\end{align}

Thus an entangled state of the spins of the two masses is obtained. The entanglement witness ${\cal W}=|\langle \sigma_x^{(1)} \otimes \sigma_z^{(2)} \rangle + \langle \sigma_y^{(1)} \otimes \sigma_y^{(2)} \rangle|$ is calculated by measuring the spin correlations in two complementary bases. If ${\cal W}>1$, the states are entangled thus the mediating gravitational field is quantised.\par

{\em Experimental setup.-}
IBM has built 5-qubit and 16-qubit quantum computers publicly available using a cloud server \cite{qgs_quantumcomputing}. Researchers around the globe, are using it to verify several quantum algorithms and simulation of quantum mechanical processes \cite{ qgs_sayanmanabputra,qgs_WhitfieldMP2011,qgs_SisodiaQIP2017,qgs_WoottonQST2017,qgs_BertaNJP2016,qgs_DeffnerHel2017,qgs_HuffmanPRA2017,qgs_AlsinaPRA2016,qgs_GarciaarXiv2017,qgs_DasarXiv2017,qgs_BKB1QIP2017,qgs_YalcinkayaPRA2017,qgs_GhosharXiv2017,qgs_KandalaNAT2017,qgs_Solano2arXiv2017,qgs_SchuldEPL2017,qgs_SisodiaPLA2017}. We use IBM's 5-qubit quantum computer `ibmqx4' to simulate the scheme. The experimental architecture of the chip `ibmqx4' is given in Table \ref{qgs_tab1}, where gate and readout errors, coherence and relaxation time of qubits are presented.
\onecolumngrid

\begin{table}[H]
\centering
\begin{tabular}{ c c c c c  }
\hline
\hline
Qubits &  $T^{\dagger}_{1}$ ($\mu s$) & $T^{\ddagger}_{2}$ ($\mu s$) & GE$^{||}$ (10$^{-3}$) & RE$^{\perp}$ $(10^{-2})$\\
\hline
q[0] & 50.81 & 14.70 & 0.86 & 4.80 \\
q[1] & 50.00 & 64.60 & 1.46 & 5.30 \\
q[2] & 47.90 & 45.00 & 1.29 & 9.80 \\ 
q[3] & 37.40 & 15.10 & 3.44 & 5.70 \\
%Q4 & 6.52698 & 5.1824 & -332.5 & 458 & 49.5 & 19.2\\
\hline
\hline
\end{tabular}\\
$\dagger$ Coherence time, $\ddagger$ Relaxation time, $||$ Gate Error, $\perp$ Readout Error. 
\caption{\textbf{The table details the experimental parameters of the quantum chip `ibmqx4'.}}
\label{qgs_tab1}
\end{table} 

\begin{figure}[H]
   \begin{center}
   \includegraphics[scale=0.56]{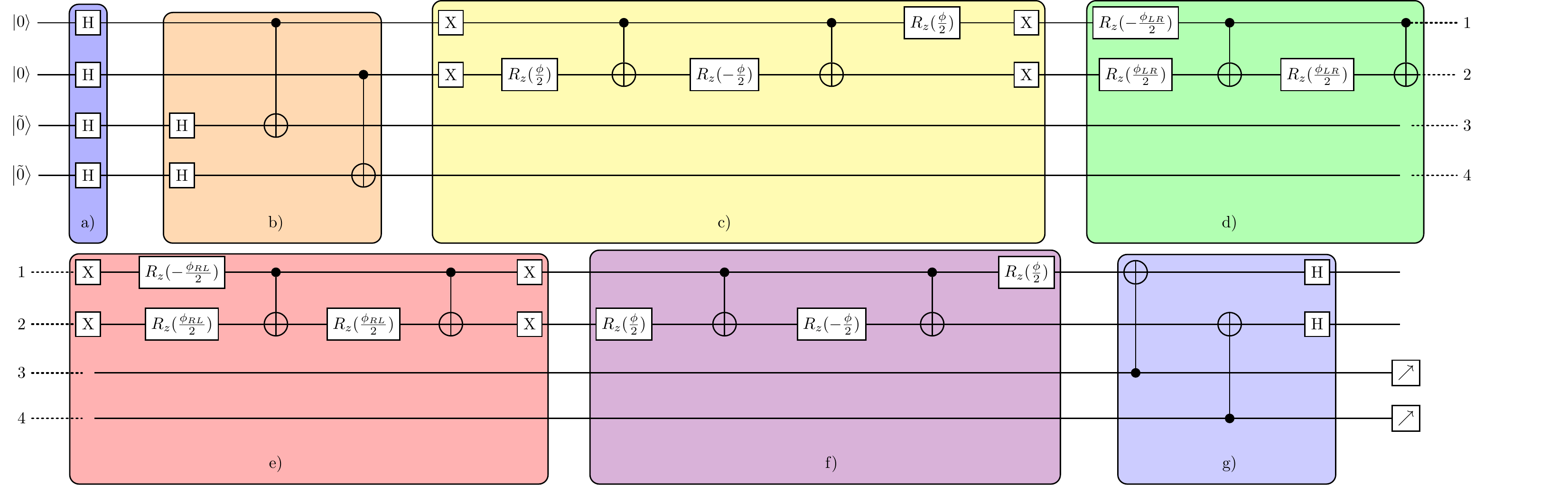}
   \end{center}
    \caption{\textbf{Quantum circuit for simulating the scheme.} $|0\rangle$ and $|\tilde{0}\rangle$ denotes the mass quantum states and the spin quantum states respectively. \textbf{Step a:} we create superposition of both mass and spin quantum states using Hadamard gates. \textbf{Step b:} we entangle the mass qubit with the corresponding spin qubit of either of the test masses. \textbf{Step c, d, e, f:} we introduce the gravitational phase to the system via `LOCC'. \textbf{Step g:} we disentangle mass and the spin qubits and then measure the spin qubits in two complementary  bases to calculate entanglement witness ${\cal W}$. The detailed decomposition of the quantum circuit is explained in the Supplemental material \ref{sec_sup}. }
    \label{qgs_Fig1}
\end{figure}

\twocolumngrid

{\em Quantum circuit.-}
We propose a quantum circuit to implement the scheme in `ibmqx4'. We take two pair of entangled qubits each corresponding to one test mass particle where the first qubit of each pair denotes the mass eigenstate and the second one denotes the spin eigenstates.
Then we introduce the gravitational phases to the system using a set of unitary operations. Finally, these phases thus entangle the mass qubits. By measuring the spin qubits in two complementary bases, we calculate the entanglement witness ($\cal W$).

{\em Result Analysis.-}
By varying the value of $\Delta\phi_{LR}+\Delta\phi_{RL}$, we take different measurements in two complementary bases. We plot a curve between entanglement witness ($\cal W$) and $\Delta\phi_{LR}+\Delta\phi_{RL}$ to get the range of values of the phase for which $\cal W>$1 i.e. the masses are entangled. 
We find that for the range $2.9113<\Delta\phi_{LR}+\Delta\phi_{RL}<4.2647$, the masses are entangled. We set the  gravitational phase values $\phi_{RL} \sim  \frac{G m_1 m_2 \tau}{\hbar (d-\Delta x)},\,\,\,\phi_{LR} \sim  \frac{G m_1 m_2 \tau}{\hbar (d+\Delta x)},\,\,\, \phi \sim \frac{G m_1 m_2 \tau}{\hbar d}$ accordingly to achieve the desired $\Delta\phi_{LR}+\Delta\phi_{RL}$ values for the simulation.\par

We know that Casimir-Polder interaction can also introduce phases mimicking the effects of gravitational interaction \cite{qgs_CasimirPR1948}. To minimize this effect, the distance of closest approach $(d-\Delta x)$ should be $\sim200\mu m$, for which
\begin{figure}[H]
   \begin{center}
    \includegraphics[scale=0.58]{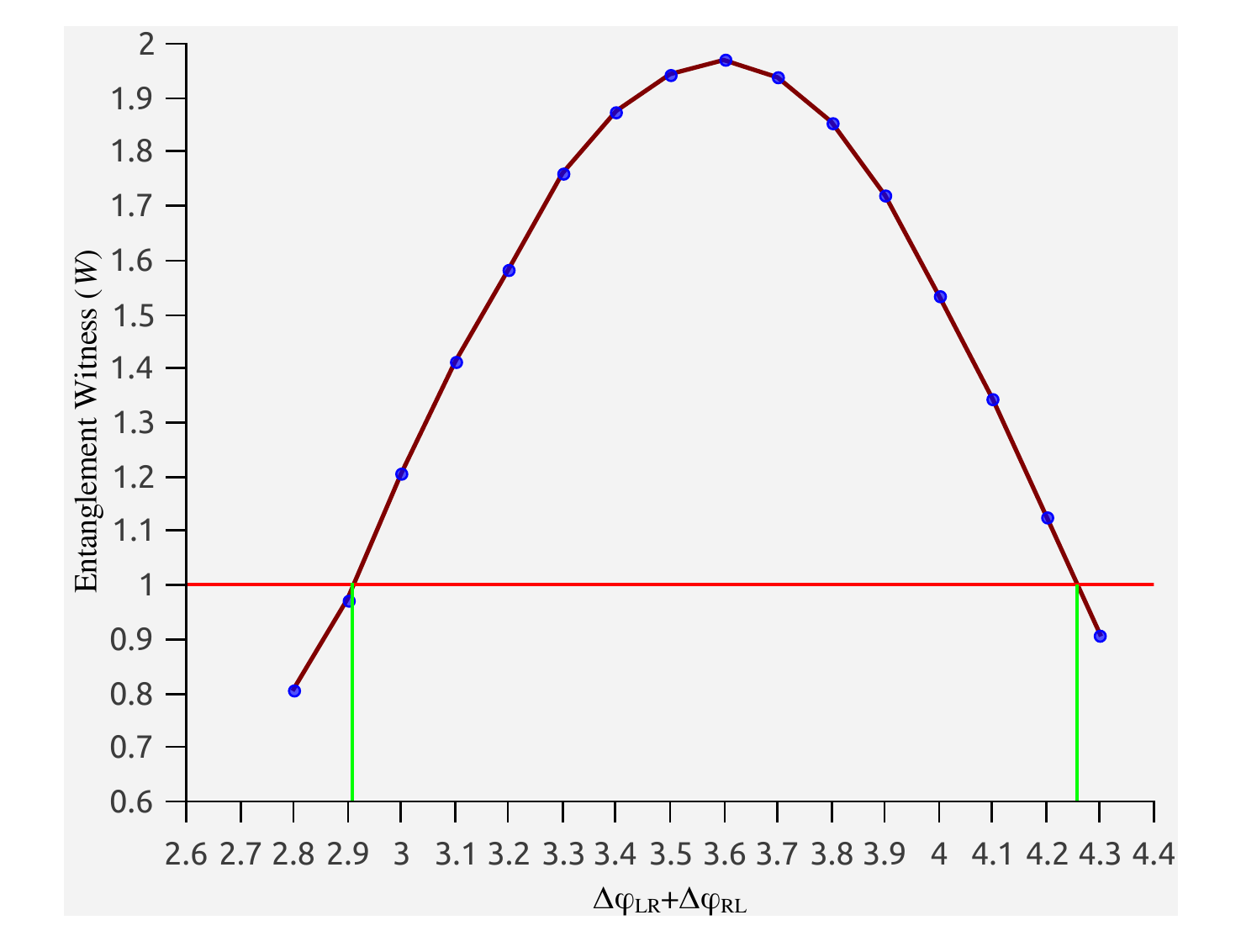}
   \end{center}
    \caption{\textbf{Entanglement Witness ($\cal W$) vs $\Delta\phi_{LR}+\Delta\phi_{RL}$ plot}. Data points in the upper half of the horizontal red line denotes entangled states. The section of the x-axis in between two vertical green lines are the corresponding $\Delta\Phi_{LR}+\Delta\phi_{RL}$ values for which the state is entangled.}
    \label{qgs_Fig2}
\end{figure}
the Casimir-Polder potential is negligible \cite{qgs_BosePRL2017}. Also, the expression $\Delta \phi_{LR} + \Delta \phi_{RL} \neq 2n\pi$ should hold.\\

As proposed by Bose \emph{et al.} \cite{qgs_BosePRL2017}, we take $m_1,m_2\sim10^{-14}$ kg, $d=450$ $ \mu$m and $\Delta x=250$ $ \mu$m for the simulation. [As interferometry in this mass scale have been proposed \cite{qgs_PinoarXiv2018}]. we can now vary the coherence time ($\tau$) to get different phases. From the quantum simulation, we observed that for $46.7439s<\tau<68.4742s$ the masses are entangled. Now, given current technology, it is difficult to ensure this much coherence time in the real experiment. Also, with long coherence time, there can be direct spin-spin dipolar interaction. \par
Alternatively, as Marletto and Vedral \cite{qgs_MarlettoPRL2017} suggested, we can take the masses as $m_1,m_2\sim10^{-12}$ kg, by using massive molecules, two split Bose condensates or two nanomechanical oscillators \cite{qgs_AsenbaumPRL2017,qgs_ChanNat2011,qgs_ArndtNat1999}. Thus, required coherence time is reduced to 0.4674s$<\tau<$0.6847s which is quite feasible using current technology.    

{\em Summary.-} To conclude, we have simulated the quantumness of gravity  using IBM quantum computer `ibmqx4'. We have designed a quantum circuit for the first time which essentially verifies the schemes proposed by Bose \emph{et al.} \cite{qgs_BosePRL2017} and Marletto and Vedral \cite{qgs_MarlettoPRL2017} in a quantum system that uses superconducting qubits and studied the entanglement characteristics. Under some specific parameters, this quantum system is analogous to quantum gravity and shows similar experimental outcomes. We observed that the degree of entanglement substantially increased and then decreased over a range of phase value. We have taken specific phase values to maximize this entanglement. Recently, there has been an argument \cite{qgs_AnastopoulosARXIV2018} that this scheme does not prove the quantumness of gravity, as  gravitational degrees of freedom used in this scheme are the pure gauge that has neither classical nor quantum physical content. So, whether gravity is a quantum entity or not still remains as an open question. Nevertheless, our attempt is to experimentally realize the scheme and give an analogue gravitational approach to it by observing a quantum system that has effects similar to that of quantum gravity. In future, these results can be improved by using a quantum computer with higher processing capacity and which can ensure a longer coherence time. If we can push beyond our current technological constraints, hopefully, we will be able to verify the quantum nature of gravity in a laboratory based experiment.
\par
{\em \textbf{Author contributions.-}} M. developed designed the quantum circuit, drew the circuit on IBM quantum experience platform, performed the experiments, interpreted and analyzed the experimental data. M. and B.K.B. contributed to the composition of the manuscript. B.K.B. has supervised the project. P.K.P. thoroughly checked and reviewed the manuscript. M. and B.K.B. completed the project under the guidance of P.K.P. 

{\em \textbf{Acknowledgements.-}} We thank Prof. Sougato Bose for useful discussions; Avinash Dash for his assistance with the experiment. M. and B.K.B. are financially supported respectively by Kishore Vaigyanik Protsahan Yojana (KVPY) Fellowship and INSPISE Fellowship from Department of Science and Technology (DST), Govt. of India. We thank Indian Institute of Science Education and Research Kolkata for providing hospitality during which this work was completed. We acknowledge the support of IBM Quantum Experience for providing access to the quantum processors. The discussions and opinions developed in this paper are only those of the authors and do not reflect the opinions of IBM or any of it's employees.

\newpage

\onecolumngrid
\newpage
\section*{Supplemental Material: A Simulational Model for Witnessing Quantum Effects of Gravity Using IBM Quantum Computer}\label{sec_sup}
\section{Introduction of gravitational phases in the system}
\begin{figure}[H]
   \begin{center}
    \includegraphics[scale=0.58]{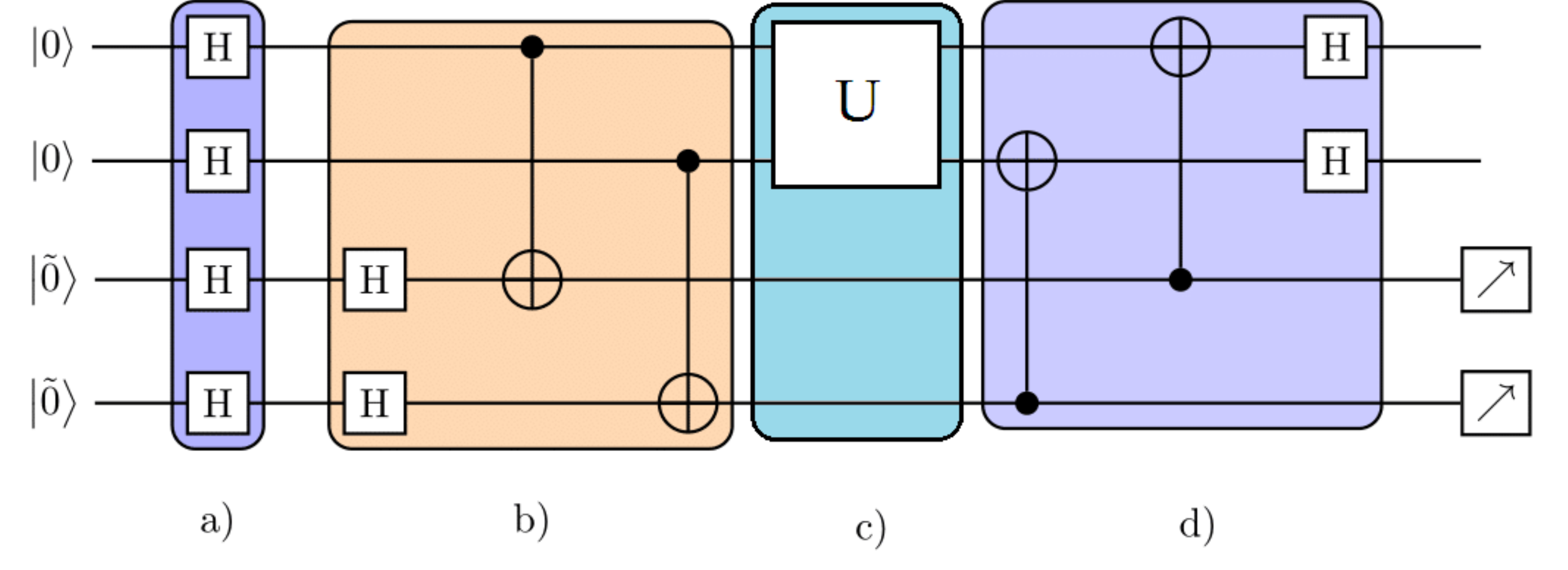}
   \end{center}
    \caption{\textbf{Proposed quantum circuit for introducing gravitational phases to the system.} Here `U' is the rotation operator corresponding to the matrix U. $|0\rangle$ and $|\tilde{0}\rangle$ denote the mass quantum states and the spin quantum states respectively. \textbf{Step a:} we create superposition of both mass and spin quantum states using Hadamard gates. \textbf{Step b:} we entangle the mass qubit with the corresponding spin qubit of either of the test masses. \textbf{Step c:} we introduce the gravitational phase to the system. \textbf{Step d:} we disentangle mass and the spin qubits and then measure the spin qubits.}
    \label{qgs_sup_Fig1}
\end{figure}
\begin{figure}[H]
   \begin{center}
    \includegraphics[scale=0.58]{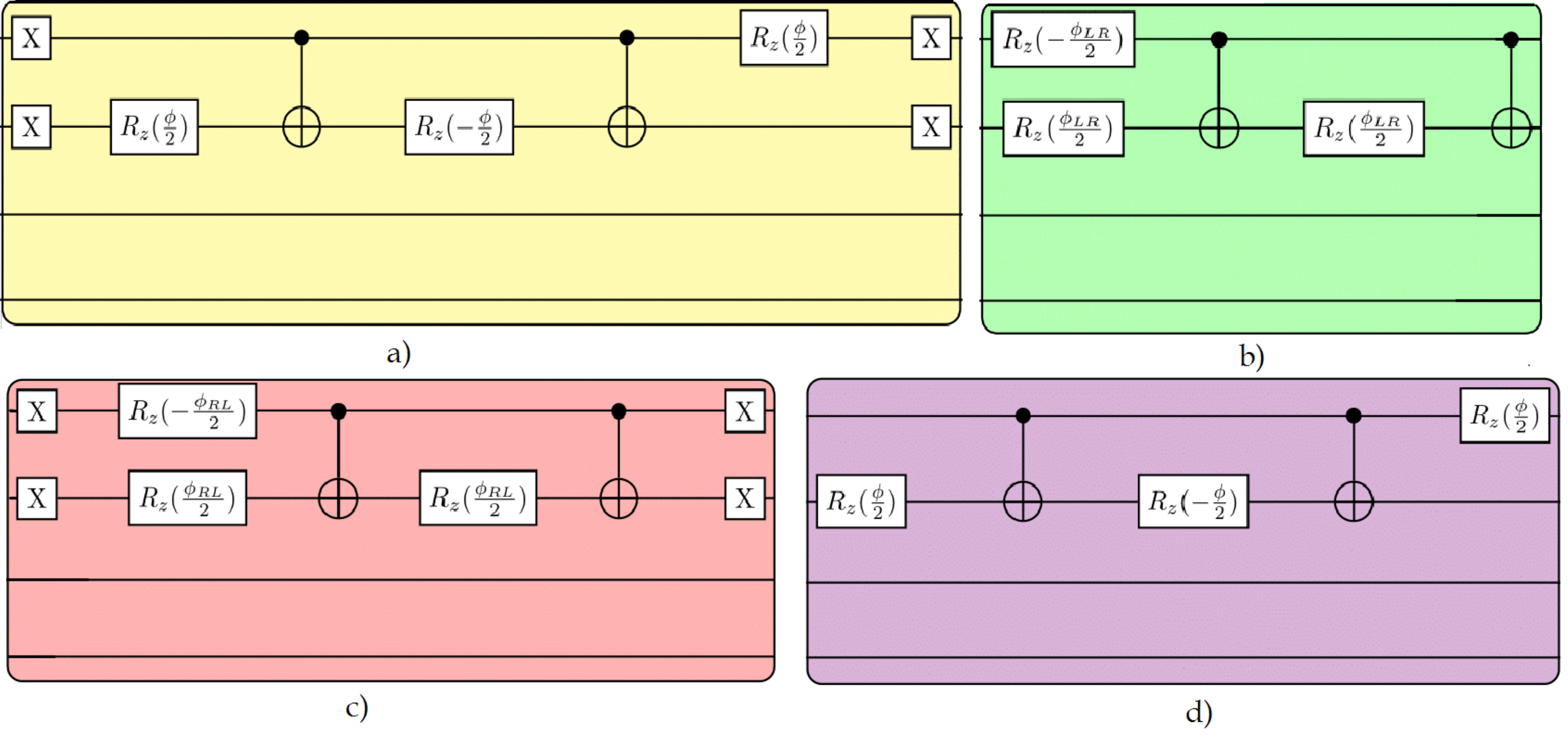}
   \end{center}
    \caption{\textbf{Circuit corresponding the four unitary matrices \cite{qgs_sup_WhitfieldMP2011}}. 
    \textbf{Case a:} circuit corresponding to $U_1^{\dagger}$, 
    \textbf{Case b:}, circuit corresponding to $U_2^{\dagger}$, 
    \textbf{Case c:}, circuit corresponding to $U_3^{\dagger}$,  
    \textbf{Case d:}, circuit corresponding to $U_4^{\dagger}$.}
    \label{qgs_sup_Fig2}
\end{figure}

We have a 4-qubit system where we need to introduce the gravitational phases in the first 2 qubits. The corresponding unitary matrix is 

\[U=
\begin{bmatrix}
e^{i\phi} &0&0&0\\
0& e^{i\phi_{LR}} &0&0\\
0&0& e^{i\phi_{RL}} &0\\
0&0&0& e^{i\phi}
\end{bmatrix}
\]
The corresponding circuit is shown in Fig. \ref{qgs_sup_Fig2}.
We need to decompose this matrix into simpler unitary matrices that can be directly applied in IBM quantum computer \cite{qgs_sup_NielsenCUP2000,qgs_sup_benenti-casatiWSP2004}.
We decompose `U' into 4 matrices such that,
\begin{align}
U_4U_3U_2U_1U=I\\
\Rightarrow U=U_4^{\dagger}U_3^{\dagger}U_2^{\dagger}U_1^{\dagger}
\end{align}
We now have decomposed matrices as,
\[U_1^{\dagger}=
\begin{bmatrix}
e^{i\phi} &0&0&0\\
0& 1 &0&0\\
0&0& 1 &0\\
0&0&0& 1
\end{bmatrix}
, U_2^{\dagger}=
\begin{bmatrix}
1 &0&0&0\\
0& e^{i\phi_{LR}} &0&0\\
0&0& 1 &0\\
0&0&0& 1
\end{bmatrix}
, U_3^{\dagger}=
\begin{bmatrix}
1 &0&0&0\\
0& 1 &0&0\\
0&0& e^{i\phi_{RL}} &0\\
0&0&0& 1
\end{bmatrix}
and\quad U_4^{\dagger}=
\begin{bmatrix}
1 &0&0&0\\
0& 1 &0&0\\
0&0& 1 &0\\
0&0&0& e^{i\phi}
\end{bmatrix}
\]

We decompose the matrix further using Hadamard, Pauli X, CNOT and Phase ($R_Z(\phi)$) gates. Thus it can finally be implemented in IBM quantum computer.
Here, $R_Z(\phi)$ is the unitary rotation operator for implementing in `ibmqx4'.
\[R_Z(\phi)=
\begin{bmatrix}
1&0\\

0& e^{i\phi}
\end{bmatrix}
\]
We measure the spin qubits in two complementary bases to calculate the entanglement witness (${\cal W})=|\langle \sigma_x^{(1)} \otimes \sigma_z^{(2)} \rangle + \langle \sigma_y^{(1)} \otimes \sigma_y^{(2)} \rangle|$. The standard measurement basis for quantum chip `ibmqx4' is $\sigma_z$. Hence for measuring in $\sigma_z$ basis we add no further gate. To measure in $\sigma_x$ basis we apply a Hadamard gate and for $\sigma_y$ basis we apply one $S^{\dagger}$ and then a Hadamard gate before measurement. 

\section{Quantum circuit and  QASM code used in `ibmqx4'}

\begin{figure}[H]
   \begin{center}
    \includegraphics[width=\linewidth]{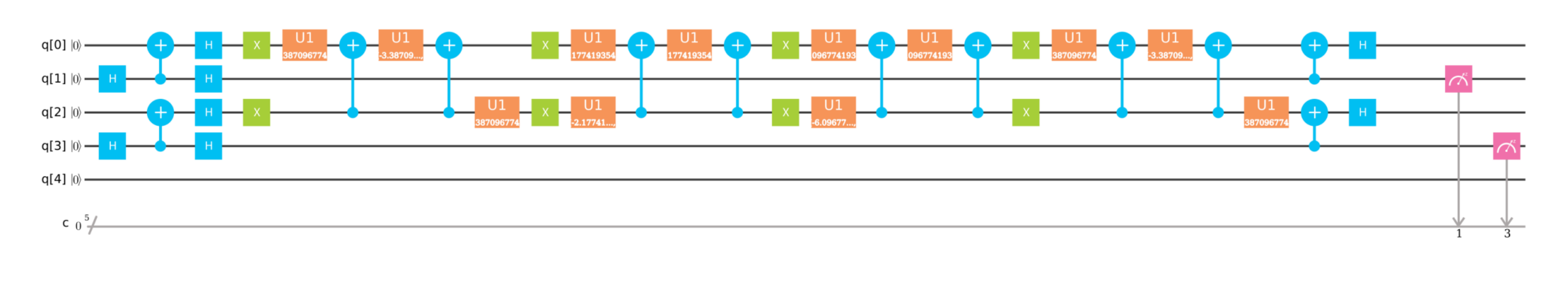}
   \end{center}
    \caption{\textbf{Circuit as implemented in IBM quantum computer `ibmqx4'.} Here `q[0]' and `q[2]' are the mass qubits and `q[1]' and `q[3]' are the respective spin qubits.}
    \label{qgs_sup_Fig3}
\end{figure}

\newpage
The QASM code of the circuit used in the simulation is presented below.
\begin{lstlisting}[language=Python]
OPENQASM 2.0;
include "qelib1.inc";
qreg q[5];
creg c[5];

#Creating equal superposition state and entangling mass qubits with corresponding spin qubits 
h q[1];
h q[3];
cx q[1],q[0];
cx q[3],q[2];
h q[0];
h q[1];
h q[2];
h q[3];

#Introducing gravitational phase to the system
x q[0];
x q[2];
u1(3.3870967742) q[0];
cx q[2],q[0];
u1(-3.3870967742) q[0];
cx q[2],q[0];
u1(3.3870967742) q[2];
x q[0];
x q[2];
u1(2.1774193549) q[0];
u1(-2.1774193549) q[2];
cx q[2],q[0];
u1(2.1774193549) q[0];
cx q[2],q[0];
x q[0];
x q[2];
u1(6.0967741935) q[0];
u1(-6.0967741935) q[2];
cx q[2],q[0];
u1(6.0967741935) q[0];
cx q[2],q[0];
x q[0];
x q[2];
u1(3.3870967742) q[0];
cx q[2],q[0];
u1(-3.3870967742) q[0];
cx q[2],q[0];
u1(3.3870967742) q[2];

#Disentangling The mass qubits from the corresponding spin qubits.
cx q[1],q[0];
cx q[3],q[2];
h q[0];
h q[2];

#Measuring the spin qubits
measure q[1] -> c[1];
measure q[3] -> c[3];
\end{lstlisting}


\begin{thebibliography}{50}


\bibitem{qgs_SinghCurrSci2015}T.~P. Singh, \href{\doibase 10.18520/v109/i12/2258-2264}{Curr. Sci. \textbf{109}, 2258 (2015)}. 

\bibitem{qgs_KhrennikovFoundPhys2017}A. Khrennikov, \href{\doibase 10.1007/s10701-017-0089-0}{Found. Phys. \textbf{47}, 1077 (2017)}. 

\bibitem{qgs_WittGAQFT2003}D. DeWitt, \textit{The Global Approach to Quantum Field Theory} International Series of Monographs on Physics (Oxford University Press, Oxford, 2003).

\bibitem{qgs_KibbleCMP1979}T.~W.~B. Kibble, \href{\doibase 10.1007/BF01225149}{Commun. Math. Phys. \textbf{65}, 189 (1979)}.

\bibitem{qgs_PolchinskiIBS2005}J. Polchinski, \textit{Introduction to the Bosonic String} (Cambridge Monographs on Mathematical Physics, Cambridge, 2005), Vol. 1.

\bibitem{qgs_KieferIMP2013}C. Kiefer, \href{\doibase 10.1155/2013/509316}{ISRN Math. Phys. \textbf{2013}, 509316 (2013)}.

\bibitem{qgs_IshamICGRG1995} C. Isham, in 14th International Conference on General Relativity and Gravitation, Florence, Italy, edited by M. Francaviglia, G. Longhi, L. Lusanna, and E. Sorace (World Scientific, Singapore, 1995), pp. 167–209.

\bibitem{qgs_BorzeszkowskiTMQG1988} H.~H. von Borzeszkowski and H.-J. Treder, \textit{The Meaning of Quantum Gravity}, Treder. Fundamental Theories of Physics Vol. 20 (D. Reidel Publishing Co., Dordrecht, The Netherlands, 1988), pp. 8–132.

\bibitem{qgs_DysonIJMP2012}F. Dyson, \href{\doibase 10.1142/S0217751X1330041X}{Int. J. Mod. Phys. A \textbf{28}, 25 (2012)}. 

\bibitem{qgs_BoughnCGG2006}S. Boughn and T. Rothman, \href{\doibase 10.1088/0264-9381/23/20/006}{Class. Quant. Grav. \textbf{23}, 5839 (2006)}. 

\bibitem{qgs_KieferAP2006}C. Kiefer, \href{\doibase 10.1002/andp.200510175}{Ann. Phys. (Amsterdam) \textbf{15}, 129 (2006)}.

\bibitem{qgs_RothmanFP2006}T. Rothman and S. Boughn, \href{\doibase 10.1007/s10701-006-9081-9}{Found. Phys. \textbf{36}, 1801 (2006)}.

\bibitem{qgs_OritiAQG2009}\textit{Approaches to Quantum Gravity}, edited by D. Oriti (Cambridge University Press, Cambridge, 2009).

\bibitem{qgs_BiswasPRL2012}T. Biswas, E. Gerwick, T. Koivisto, and A. Mazumdar, \href{\doibase 10.1103/PhysRevLett.108.031101}{Phys. Rev. Lett. \textbf{108}, 031101 (2012)}.
 
\bibitem{qgs_HoseenfelderCQG2011}S. Hossenfelder, \textit{Classical Quantum Gravity: Theory, Analysis and Applications}, edited by V. R. Frignanni (Nova Publishers, 2011), Chap. 5.
\bibitem{qgs_AshoorioonPLA2014}A. Ashoorioon, P.~S. Bhupal Dev, and A. Mazumdar, \href{\doibase 10.1142/S0217732314501636}{Mod. Phys. Lett. A \textbf{29}, 1450163 (2014)}.
\bibitem{qgs_PikovskiNatPhys2012}I. Pikovski, M.~R. Vanner, M. Aspelmeyer, M.~S. Kim, and C. Brukner, \href{\doibase 10.1038/nphys2262}{Nat. Phys. \textbf{8}, 393 (2012)}. 
\bibitem{qgs_AlbrechtPRA2014}A. Albrecht, A. Retzker, and M. B. Plenio, \href{\doibase 10.1103/PhysRevA.90.033834}{Phys. Rev. A \textbf{90}, 033834 (2014)}.

\bibitem{qgs_ShandarinPRL1995}S.~F. Shandarin, A.~L. Melott, K. McDavitt, J.~L. Pauls, and J. Tinker, \href{\doibase 10.1103/PhysRevLett.75.7}{Phys. Rev. Lett. \textbf{75}, 7 (1995)}.
\bibitem{qgs_SakharovGRG2000}A.~D. Sakharov, \href{\doibase 10.1023/A:1001947813563}{Gen. Relativ. Gravit. \textbf{32}, 365 (2000)}.
\bibitem{qgs_PenroseGRG1996}R. Penrose, \href{\doibase 10.1007/BF02105068}{Gen. Relativ. Gravit. \textbf{28}, 581 (1996)}. 

\bibitem{qgs_RosenfeldQGE1966}E. Rosenfeld, in \textit{Quantentheorie und Gravitation in Einstein-Symposium 1965, Berlin} (Akademie, Berlin, 1966); English translation Selected Papers of L. Rosenfeld, \textit{BostonStudies in the Philosophy of Science}, edited by R. S. Cohen and J. J. Stachelin (Reidel, Dordrecht, 1979), p. 599.

\bibitem{qgs_GorelikPU2005}G. Gorelik, \href{\doibase 10.1070/PU2005v048n10ABEH005820}{Phys. Usp. \textbf{48}, 1039 (2005)}.
\bibitem{qgs_BronsteinPZS1936}M.~P. Bronstein, Phys. Z. Sowjetunion \textbf{9.2–3}, 140 (1936).
\bibitem{qgs_FeynmanCHCP1957}R. Feynman, in \textit{Chapel Hill Conference Proceedings}, 1957.

\bibitem{qgs_BahramiarXiv2015}M. Bahrami, A. Bassi, S. McMillen, M. Paternostro, and H. Ulbricht, \href{https://arxiv.org/abs/1507.05733}{arXiv:1507.05733}.
\bibitem{qgs_PagePRL1981}D.~N. Page and C.~D. Geilker, \href{\doibase 10.1103/PhysRevLett.47.979}{Phys. Rev. Lett. \textbf{47}, 979 (1981)}.
\bibitem{qgs_AnastopoulosCGG2015}C. Anastopoulos and B.-L. Hu, \href{\doibase 10.1088/0264-9381/32/16/165022}{Classical Quantum Gravity \textbf{32}, 165022 (2015)}.
\bibitem{qgs_DerakhshaniJPCS2016}M. Derakhshani, C. Anastopoulos, and B.-L. Hu, \href{}{J. Phys. Conf. Ser. \textbf{701}, 012015 (2016)}.
\bibitem{qgs_DerakhshaniarXiv2016}M. Derakhshani, \href{https://arxiv.org/abs/1609.01711}{arXiv:1609.01711}.
\bibitem{qgs_MarlettoNat2017}C. Marletto and V. Vedral, \href{\doibase 10.1038/547156a}{Nature (London) \textbf{547}, 156 (2017)}.

\bibitem{qgs_MarlettoarXiv2018}C. Marletto and V. Vedral, \href{https://arxiv.org/abs/1803.09124}{arXiv:1803.09124}.

\bibitem{qgs_DiosiPLA1987}L. Diosi, \href{\doibase 10.1016/0375-9601(87)90681-5}{Phys. Lett. A \textbf{120}, 377 (1987)}.
\bibitem{qgs_KharolihazyNC1966}F. Kharolihazy, \href{\doibase 10.1007/BF02717926}{Nuovo Cimento A \textbf{42}, 390 (1966)}.

\bibitem{qgs_ColellaPRL1975}R. Colella, A.~W. Overhauser, and S.~A. Werner, \href{\doibase 10.1103/PhysRevLett.34.1472}{Phys. Rev. Lett. \textbf{34}, 1472 (1975)}.

\bibitem{qgs_MargalitSci2015}Y. Margalit, Z. Zhou, S. Machluf, D. Rohrlich, Y. Japha, and R. Folman, \href{\doibase 10.1126/science.aac6498}{Science \textbf{349}, 1205 (2015)}.
\bibitem{qgs_PikovskiNatPhys2015}I. Pikovski, M. Zych, F. Costa, and \v{C}. Brukner, \href{\doibase 10.1038/nphys3366}{Nat. Phys. \textbf{11}, 668 (2015)}.
\bibitem{qgs_AhluwaliaPRD1998}D.~V. Ahluwalia and C. Burgard, \href{\doibase 10.1103/PhysRevD.57.4724}{Phys. Rev. D \textbf{57}, 4724 (1998)}.


\bibitem{qgs_HorodeckiRMP2009}R. Horodecki, P. Horodecki, M. Horodecki, and K. Horodecki, \href{\doibase 10.1103/RevModPhys.81.865}{Rev. Mod. Phys. \textbf{81}, 865 (2009)}.

\bibitem{qgs_SchrodingerPCPS1935}E. Schr$\ddot{o}$dinger, \href{\doibase 10.1017/S0305004100013554}{Proc. Cambridge Philos. Soc. \textbf{31}, 555 (1935)}.

\bibitem{qgs_BosePRL2017}S. Bose, A. Majumdar, G.~W. Morley, H. Ulbricht, M. Toro\v{s}, M. Paternostro, A.~A. Geraci, P.~F. Barker, M.~S. Kim, and G. Milburn, \href{\doibase 10.1103/PhysRevLett.119.240401}{Phys. Rev. Lett. \textbf{119}, 240401 (2017)}.




\bibitem{qgs_MarlettoPRL2017}C. Marletto and V. Vedral, \href{\doibase 10.1103/PhysRevLett.119.240402}{Phys. Rev. Lett. \textbf{119}, 240402 (2017)}.

\bibitem{qgs_CasimirPR1948}H.~B.~G. Casimir and D. Polder, \href{10.1103/PhysRev.73.360}{Phys. Rev. \textbf{73}, 360 (1948)}.


\bibitem{qgs_quantumcomputing}IBM Quantum Experience.\href{http://research.ibm.com/ibm-q/}{ http://research.ibm.com/ibm-q/}

\bibitem{qgs_WhitfieldMP2011}J.~D. Whitfield, J. Biamonte, A. Aspuru-Guzik, \href{\doibase 10.1080/00268976.2011.552441}{Molecul. Phys. \textbf{109}, 735 (2011)}. 

\bibitem{qgs_sayanmanabputra}S. Gangopadhyay, Manabputra, B.~K. Behera, and P.~K. Panigrahi, \href{\doibase 10.1007/s11128-018-1932-8}{Quantum Inf. Process. \textbf{17}, 160 (2018)}.
 
\bibitem{qgs_SisodiaQIP2017}M. Sisodia, A. Shukla, K. Thapliyal, and A. Pathak, \href{\doibase 10.1007/s11128-017-1744-2}{Quantum Inf. Process. \textbf{16}, 292 (2017)}.    

\bibitem{qgs_WoottonQST2017}J.~R. Wootton, \href{\doibase 10.1088/2058-9565/aa5c73}{Quantum Sci. Technol. \textbf{2}, 015006 (2017)}. 

\bibitem{qgs_BertaNJP2016}M. Berta, S. Wehner, and M.~M. Wilde, \href{\doibase 10.1088/1367-2630/18/7/073004}{New J. Phys. \textbf{18}, 073004 (2016)}. 

\bibitem{qgs_DeffnerHel2017}S. Deffner, \href{\doibase 10.1016/j.heliyon.2017.e00444}{Heliyon \textbf{3}, e00444 (2016)}. 

\bibitem{qgs_HuffmanPRA2017}E. Huffman and A. Mizel, \href{\doibase 10.1103/PhysRevA.95.032131}{Phys. Rev. A \textbf{95}, 032131 (2017)}. 

\bibitem{qgs_AlsinaPRA2016}D. Alsina and J.~I. Latorre, \href{\doibase 10.1103/PhysRevA.94.012314}{Phys. Rev. A \textbf{94}, 012314 (2016)}.

\bibitem{qgs_GarciaarXiv2017}D. Garc\'{i}a-Mart\'{i}n and G. Sierra, \href{https://arxiv.org/pdf/1712.05642.pdf}{arXiv:1712.05642}.

\bibitem{qgs_DasarXiv2017}S. Das and G. Paul, \href{https://arxiv.org/abs/1712.04925}{arXiv:1712.04925}. 

\bibitem{qgs_BKB1QIP2017}B.~K. Behera, A. Banerjee, and P.~K. Panigrahi, \href{\doibase 10.1007/s11128-017-1762-0}{Quantum Inf. Process. \textbf{16}, 312 (2017)}.

\bibitem{qgs_YalcinkayaPRA2017}\.{I} Yal\c{c}inkaya and Z. Gedik, \href{\doibase 10.1103/PhysRevA.96.062339}{Phys. Rev. A \textbf{96}, 062339 (2017)}. 

\bibitem{qgs_GhosharXiv2017}D. Ghosh, P. Agarwal, P. Pandey, B.~K. Behera, and P.~K. Panigrahi, \href{\doibase 10.1007/s11128-018-1920-z}{Quantum Inf. Process. \textbf{17}, 153 (2018)}.

\bibitem{qgs_KandalaNAT2017}A. Kandala, A. Mezzacapo, K. Temme, M. Takita, M. Brink, J.~M. Chow, and J.~M. Gambetta, \href{\doibase 10.1038/nature23879}{Nature \textbf{549}, 242 (2017)}. 

\bibitem{qgs_Solano2arXiv2017}U. Alvarez-Rodriguez, M. Sanz, L. Lamata, and E. Solano, \href{https://arxiv.org/abs/1711.09442}{	arXiv:1711.09442}.

\bibitem{qgs_SchuldEPL2017}M. Schuld, M. Fingerhuth, and F. Petruccione, \href{\doibase 10.1209/0295-5075/119/60002}{Europhys. Lett. \textbf{119}, 60002 (2017)}. 

\bibitem{qgs_SisodiaPLA2017}M. Sisodia, A. Shukla, and A. Pathak, \href{\doibase 10.1016/j.physleta.2017.09.050}{Phys. Lett. A \textbf{381}, 3860 (2017)}.  


\bibitem{qgs_PinoarXiv2018}H. Pino, J. Prat-Camps, K. Sinha, B.~P. Venkatesh, O. Romero-Isart, \href{https://arxiv.org/abs/1603.01553v2}{arXiv:1603.01553}. 

\bibitem{qgs_ArndtNat1999} M. Arndt, O. Nairz, J. Vos-Andreae, C. Keller, G. van der Zouw, and A. Zeilinger, \href{\doibase 10.1038/44348}{Nature (London) \textbf{401}, 680 (1999)}.

\bibitem{qgs_ChanNat2011} J. Chan et al., \href{\doibase 10.1038/nature10461}{Nature (London) \textbf{89}, 478 (2011)}.

\bibitem{qgs_AsenbaumPRL2017} P. Asenbaum, C. Overstreet, T. Kovachy, D.~D. Brown, J.~M. Hogan, and M.~A. Kasevich, \href{\doibase 10.1103/PhysRevLett.118.183602}{Phys. Rev. Lett. \textbf{118}, 183602 (2017)}.
\bibitem{qgs_AnastopoulosARXIV2018}C. Anastopoulos and B.-L. Hu, \href{https://arxiv.org/abs/1804.11315}{arXiv: 1804.11315}.
 \bibliography{\jobname}
\end{thebibliography}

\begin{thebibliography}{}
\bibitem{qgs_sup_NielsenCUP2000}M. A. Nielsen and I. L. Chuang, \textit{Quantum Computation and Quantum Information} (Cambridge University Press, Cambridge, U.K., 2000)
\bibitem{qgs_sup_WhitfieldMP2011}J.~D. Whitfield, J. Biamonte, A. Aspuru-Guzik, \href{\doibase 10.1080/00268976.2011.552441}{Molecul. Phys. \textbf{109}, 735 (2011)}. 
\bibitem{qgs_sup_benenti-casatiWSP2004} G Benenti, G Casati and G Strini, \textit{Principles of Quantum Computation
and Information} (World Scientific Publishing, Singapore, 2004)
\end{thebibliography}
\end{document}